\documentclass[fleqn , usenatbib]{mnras}

\usepackage{newtxtext , newtxmath}

\usepackage[T1]{fontenc}

\DeclareRobustCommand{\VAN}[3]{#2}
\let\VANthebibliography\thebibliography
\def\thebibliography{\DeclareRobustCommand{\VAN}[3]{##3}\VANthebibliography}

\usepackage{graphicx}	
\usepackage{amsmath}	
\usepackage{multicol}

\usepackage[dvipsnames]{xcolor}
\usepackage{amsmath}

\newcommand{\mass}{$M_2~=~0.0993~\pm 0.0033~\,\mathrm{M_{\sun}}$}
\newcommand{\radius}{$R_2 =~0.1250~\pm0.0016~\,\mathrm{R_{\sun}}$}
\newcommand{\period}{$P =~44.92471~\pm0.00025~\,\rm{\ day}$}
\newcommand{\loggsec}{$\log g_2 = 5.222 \pm 0.0135$}
\newcommand{\massfunction}{$f_m = 0.0008127\pm0.0000028~\,\mathrm{M_{\sun}}$}
\newcommand{\eccentricity}{$e = 0.56733 \pm 0.00027$}
\newcommand{\massratio}{$q = 0.0995 \pm 0.0019$}

\title[EBLM XII - J2114-39]{The EBLM Project XII.  An eccentric, long-period eclipsing binary with a companion near the hydrogen-burning limit}
\author[Y. T. Davis et al.]
{Yasmin T. Davis $^{1}$, 
Amaury H.M.J. Triaud$^{1}$,
Alix V. Freckelton$^{1}$,
Annelies Mortier$^{1}$,
Daniel Sebastian$^{1}$,
\newauthor
Thomas Baycroft$^{1}$,
Rafael Brahm$^{2,3,4}$,
Georgina Dransfield$^{1}$,
Alison Duck$^{5}$,
Thomas Henning$^{6}$,
\newauthor
Melissa J. Hobson$^{6,3}$,
Andr\'es Jord\'an$^{2,3,4}$,
Vedad Kunovac$^{7,8}$
David V. Martin$^{5,13}$,
Pierre F.L. Maxted$^{9}$,
\newauthor
Lalitha Sairam$^{1,11}$,
Matthew R. Standing$^{10, 16}$,
Matthew I. Swayne$^{9}$,
Trifon Trifonov$^{6,14,15}$,
St\'ephane Udry$^{12}$
\\
$^{1}$School of Physics and Astronomy, University of Birmingham, Edgbaston, Birmingham B15 2TT, UK\\
$^{2}$Facultad de Ingenier\'ia y Ciencias, Universidad Adolfo Ib\'a\~nez, Av. Diagonal las Torres 2640, Pe\~{n}alol\'{e}n, Santiago, Chile\\
$^{3}$Millennium Institute for Astrophysics, Chile\\
$^{4}$Data Observatory Foundation, Chile\\
$^{5}$Department of Astronomy, The Ohio State University, Columbus, OH 43210, USA\\
$^{6}$Max-Planck-Institut für Astronomie, Königstuhl 17, 69117 Heidelberg, Germany\\
$^{7}$Department of Physics, University of Warwick, Gibbet Hill Road, Coventry CV4 7AL, UK\\
$^{8}$Centre for Exoplanets and Habitability, University of Warwick, Gibbet Hill Road, Coventry CV4 7AL, UK\\
$^{9}$Astrophysics Group, Keele University, Keele, Staffordshire, ST5 5BG, UK\\
$^{10}$School of Physical Sciences, The Open University, Walton Hall, Milton Keynes, MK7 6AA, UK\\
$^{11}$Institute of Astronomy, University of Cambridge, Madingley road, Cambridge CB3 0HA\\
$^{12}$Observatoire Astronomique de l'Universit\'e de Gen\`eve, Chemin Pegasi 51b, CH-1290 Versoix, Switzerland\\
$^{13}$NASA Sagan Fellow\\
$^{14}$Department of Astronomy, Sofia University ``St Kliment Ohridski'', 5 James Bourchier Blvd, BG-1164 Sofia, Bulgaria\\
$^{15}$Zentrum f\"ur Astronomie der Universt\"at Heidelberg, Landessternwarte, K\"onigstuhl 12, 69117 Heidelberg, Germany\\
$^{16}$European Space Agency (ESA), European Space Astronomy Centre (ESAC), Camino Bajo del Castillo s/n, 28692 Villanueva de la Cañada, Madrid, Spain\\
}

\date{Accepted XXX. Received YYY; in original form ZZZ}

\pubyear{2023}

\begin{document}
\label{firstpage}
\pagerange{\pageref{firstpage}--\pageref{lastpage}}
\maketitle

\begin{abstract}
In the hunt for Earth-like exoplanets it is crucial to have reliable host star parameters, as they have a direct impact on the accuracy and precision of the inferred parameters for any discovered exoplanet. For stars with masses between 0.35 and 0.5 ${\rm M_{\sun}}$ an unexplained radius inflation is observed relative to typical stellar models. However, for fully convective objects with a mass below 0.35 ${\rm M_{\sun}}$ it is not known whether this radius inflation is present as there are fewer objects with accurate measurements in this regime. Low-mass eclipsing binaries present a unique opportunity to determine empirical masses and radii for these low-mass stars. Here we report on such a star, EBLM J2114-39\,B. We have used HARPS and FEROS radial-velocities and \textit{TESS} photometry to perform a joint fit of the data, and produce one of the most precise estimates of a very low mass star's parameters. Using a precise and accurate radius for the primary star using {\it Gaia} DR3 data, we determine J2114-39 to be a $M_1 = 0.998 \pm 0.052$~${\rm M_{\sun}}$ primary star hosting a fully convective secondary with mass \mass, which lies in a poorly populated region of parameter space. With a radius \radius, similar to TRAPPIST-1, we see no significant evidence of radius inflation in this system when compared to stellar evolution models. We speculate that stellar models in the regime where radius inflation is observed might be affected by how convective overshooting is treated.

\end{abstract}




\section{Introduction}

The key to estimating any exoplanet's physical properties is to first determine its host star's parameters accurately. Typically, host star parameters are determined by finding the closest fit between the stellar observables and stellar models \citep[such as][]{baraffe2015, Dartmouth_models, Fernandes_lateM_model}. Thus, any biases or missing physics present in stellar models will lead to biased stellar and planetary parameter estimates. One issue of particular concern is the M-dwarf radius inflation problem where M-dwarfs with masses in the range $0.3$ to $0.5~\rm M_\odot$ appear to have radii a few percent larger than predicted by typical stellar models \citep{Torres2002,Lopez2007,Feiden2012}. Additionally, M-dwarf effective temperatures appear to fall below predictions \citep{Parsons2018}. It has been claimed that these factors compensate for each other such that the luminosity of M-dwarfs is accurately predicted by stellar models \citep{2007Torres}, but recent results have cast doubt on this claim \citep{2023Swayne}.

Although M-dwarf stars are the most common type of star in our galaxy \citep{Kroupa2001,Chabrier2003,Henry2006}, they are not as well-studied as solar-type stars. The Detached Eclipsing Binary Catalogue (DEBCat)\footnote{\url{https://www.astro.keele.ac.uk/jkt/debcat/}} archive reports only one set of double-lined eclipsing binaries with either component below $0.2~\rm M_\odot$ \citep{DEBcat, Casewell2018}, meaning that empirical mass-radius relations, determined from detached, double-lined, eclipsing binaries typically do not include fully convective M-dwarfs \citep[$\leq 0.35~\rm M_\odot$;][]{Torres2010,Moya2018}. One of the main objectives of the EBLM project \citep[Eclipsing Binaries - Low Mass;][]{EBLMI} is to measure accurate masses and radii for a sufficiently large sample of very low-mass M-dwarfs such that an empirical mass-radius-metallicity relation can be calibrated for the lower main-sequence.

Due to their low masses and low brightnesses, M-dwarf stars can be difficult to study. However, as an eclipsing secondary star they can be investigated using the same methods routinely implemented for exoplanets. For this reason, the EBLM project \citep{EBLMI} can address the low number of studied M-dwarfs and begin to fill the poorly populated region of parameter space in the mass-radius diagram for stars with masses $< 0.35~\rm M_\odot$. Populating this space is of particular importance for the study of exoplanets as low-mass stars approaching the hydrogen burning limit are ideal candidates for the detection of temperate Earth-sized planets \citep{Nutzman2008,Rodler2014,Gillon2017,Triaud2021,Delrez2022}.

Results from the EBLM survey have in the past shown that fully convective M-dwarf radii are correlated with their primary host star's metallicity, with very little evidence for a radius inflation problem \citep{Boetticher2019}. Radius inflation for high-mass M-dwarfs has been explained as the result of stellar magnetic activity whereby close-in binaries are synchronised, increasing the dynamo effect \citep[e.g.][]{Feiden2013}. If true, one expects a reduction of inflation with increasing orbital period, but results from the EBLM project show no relation between radius and orbital separation \citep{Boetticher2019, 2023Swayne}.

Here, we present a new system from the EBLM sample, EBLM~J2114-39 (hereafter J2114-39). In the past, we published radial-velocities (RVs) modelled together with eclipses from ground-based telescopes \citep[e.g.][]{EBLMI,Boetticher2019}. More recently, we combined radial-velocities with photometry from two satellites {\it CHEOPS} and {\it TESS} \citep{Swayne2021,Sebastian2023, 2023Swayne}. From now on, we will publish the rest of the EBLM systems combining HARPS and/or SOPHIE radial-velocities (as well as any available supplementary data), with photometry from {\it TESS}. This is the first such paper in the series.
In Section~\ref{sec:obs} we describe the observations collected. Section~\ref{sec:spec} details how we estimate parameters for the primary star, and the modelling of all data is in Section~\ref{sec:model}. Section~\ref{sec:adjustment} explains how we ensure we extract the most accurate parameters for the secondary star. We discuss the results and conclude in Section~\ref{sec:conclusion}.

\section{Observations}\label{sec:obs}

For our analysis of J2114-39 (other designations: TIC~159730525, 2MASS~J21143061-3918506, Gaia DR3~6583420566747746176) we used photometric data collected by {\it TESS} \citep[Transiting Exoplanet Survey Satellite;][]{TESS_inst}, alongside spectra from the HARPS \citep{HARPS_inst} and FEROS \citep{FEROS_inst} instruments. This is a single lined binary (SB1) with an eclipsing M-dwarf on a \period{} period orbiting a G-type primary. The J2114-39 system is located at $\alpha = 21^h14'30.61''$ and $\delta = -39^{\circ}18'50.6''$ on the sky with $V_{\rm mag} = 11.1$. J2114-39 was not originally included in the EBLM catalogue \citep[and is thus missing from ][]{Triaud2017} as it was identified as a likely long-period transiting gas giant, but FEROS observations quickly revealed the companion to be too massive and therefore non-planetary. The star was then added to our observing programme to be monitored by HARPS to seek circumbinary exoplanets in the context of the BEBOP survey \citep[Binaries Escorted By Orbiting Planets;][]{Martin2019,Standing2023}.

\begin{figure}
    \centering
    \includegraphics[width=0.49\textwidth]{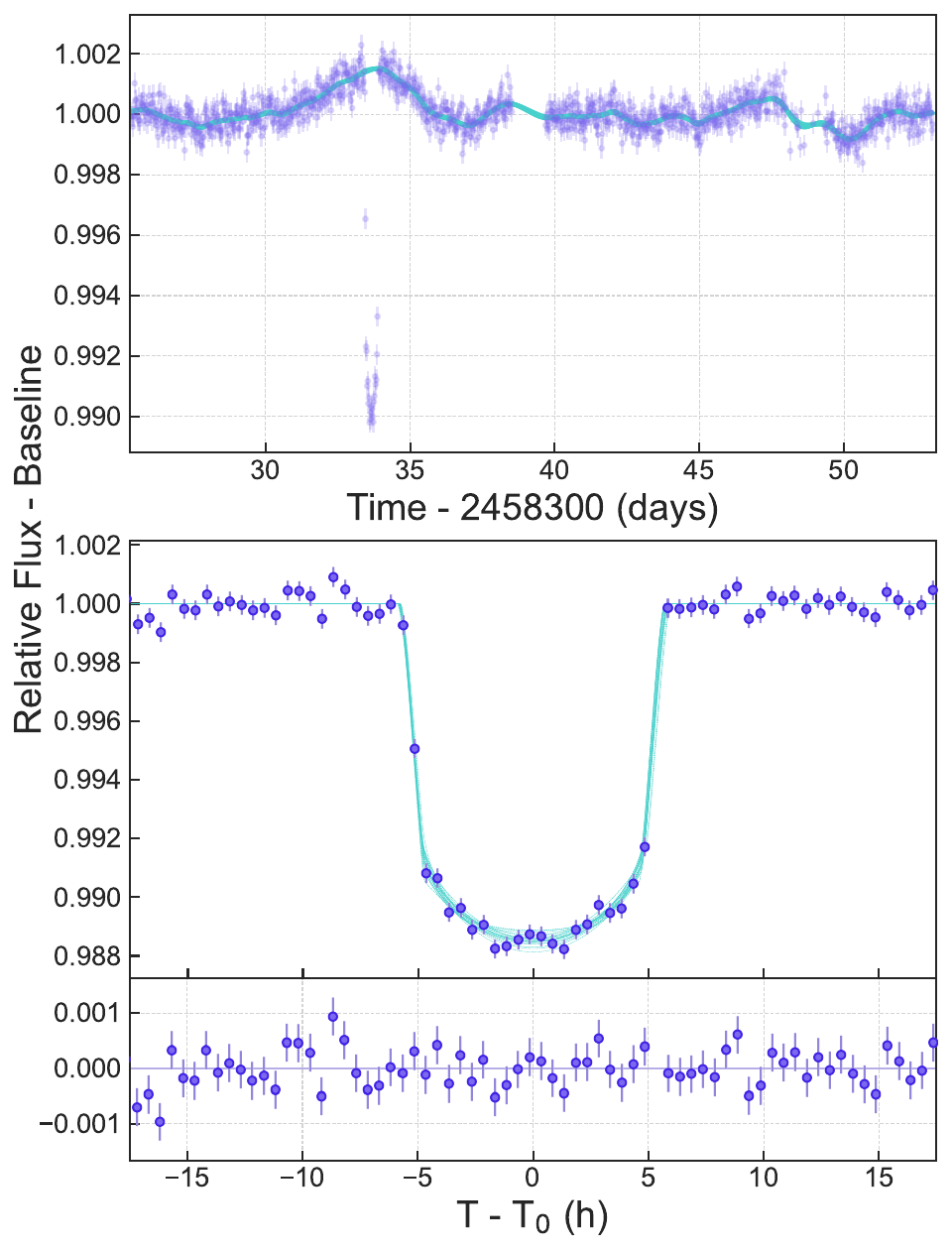}

    \caption{
    Top panel: Raw {\it TESS} data in purple with 20 random samples of the GP from the MCMC modelling the additional noise in blue. Middle panel: The phase-shifted eclipse of EBLM~J2114-39, data shown in purple after the signal modelled by the GP has been removed. Twenty random samples from the MCMC run are displayed in teal colour. Bottom panel: The residuals to the best fit model.}
    \label{fig:phased_light_curve}
\end{figure}

The observations of J2114-39 with {\it TESS} were accessed using the Mikulski Archive for Space Telescopes (MAST) web service\footnote{\url{https://mast.stsci.edu}}
and downloaded using the \texttt{lightkurve} software package \citep{lightkurve}.
The target was observed in Sector~1 where one eclipse is clearly present and Sector~28 which only contains out-of-eclipse data. The Sector 28 data is not included as it added no additional information to our fit. The phase-shifted lightcurve can be seen in Fig.~\ref{fig:phased_light_curve}. Due to the long orbital period compared to the $\sim$27 day {\it TESS} sector length, the secondary eclipse falls outside of the sectors observational windows. This analysis uses the 30-min cadence data products from the TESS-SPOC authors \citep{TESS_SPOC_pipeline} which was processed with the Presearch Data Conditioning Simple Aperture Photometry (PDCSAP) algorithm \citep{Stumpe2012,Smith2012} to remove systematic trends caused by instrumental noise. 

\begin{figure}
    \centering
    \includegraphics[width=0.49\textwidth]{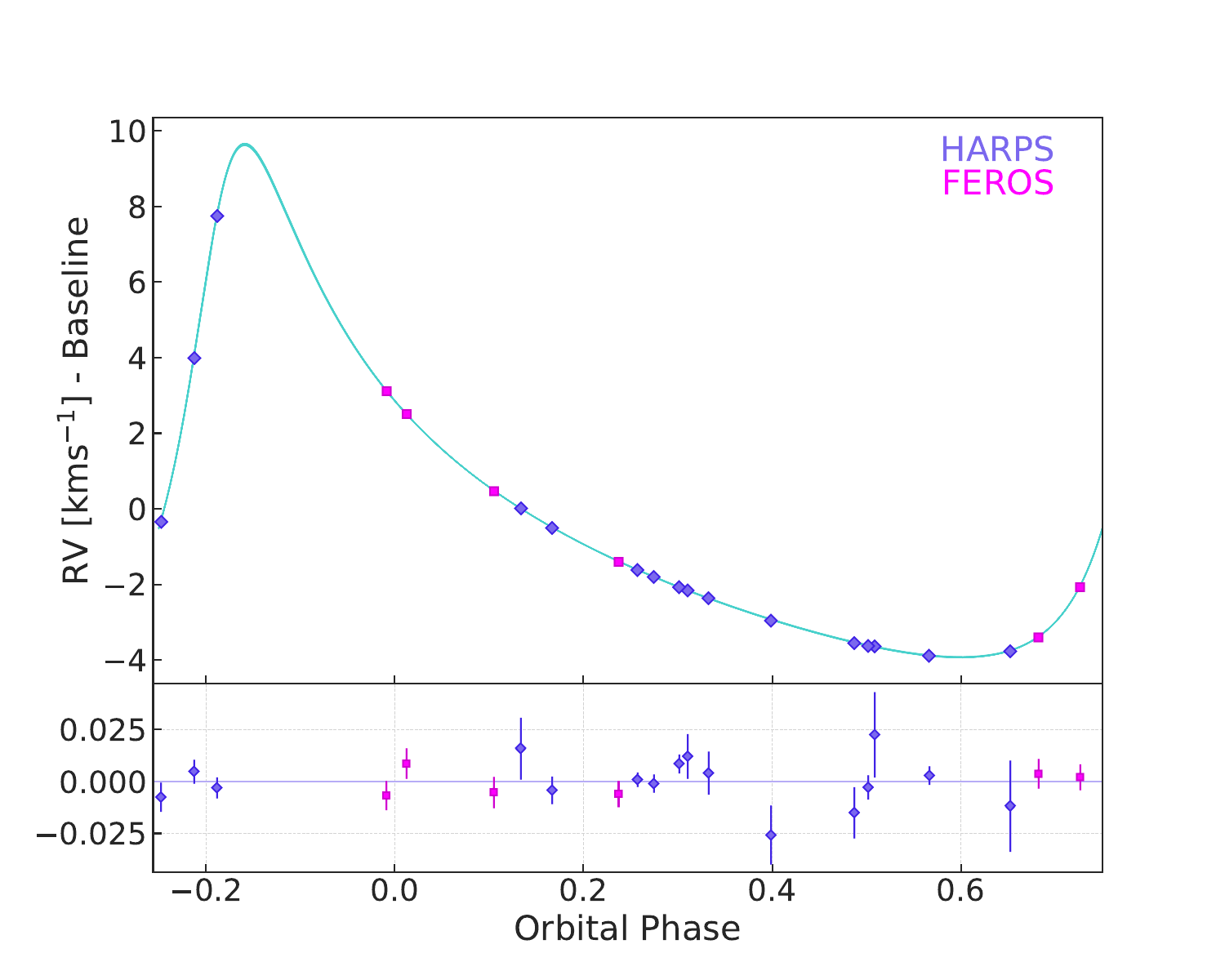}
    \caption{
    Phased radial-velocity measurements for EBLM~J2114-39 from the HARPS (in purple) and FEROS (in pink) spectrographs. A sample of 20 random models from the MCMC are shown in teal colour. The residuals to best fit model are shown in the lower panel.}
    \label{fig:phased_rv}
\end{figure}

We have six recorded radial-velocity measurements of J2114-39 with FEROS, mounted on the 2.2m telescope at La Silla Observatory between 2019 Sep 12 and Nov 2019 Nov 28, and 16 radial-velocity measurements with the HARPS spectrograph, mounted on the ESO 3.6m telescope also at La Silla Observatory between 2020 Nov 18 and 2022 Nov 14. The FEROS observations were performed in the context of the Warm gIaNts with tEss collaboration \citep[WINE,][]{hobson:2021,trifonov:2023,brahm:2023}. The radial velocities of the primary star measured from these spectra using cross-correlation against a numerical mask based on the solar spectrum are given in Tables \ref{tab:harps_obvs} and \ref{tab:feros_obvs} for the HARPS and FEROS observations respectively, and the phased radial-velocity curve can be seen in Fig.~\ref{fig:phased_rv}.

These six FEROS measurements were acquired with the simultaneous wavelength calibration technique where the second fibre is illuminated by the ThAr lamp to trace instrumental radial velocity drifts during the science exposure. We adopted an exposure time of 900\,s, which produced spectra with a typical signal-to-noise ratio of 100 per resolution element. FEROS data were processed with the \texttt{ceres} \citep{ceres} pipeline, which generates optimally extracted, wavelength calibrated, and drift corrected spectra from the raw science images. Then we computed the radial velocities with the cross-correlation technique by using a G2 binary mask as template, where a gaussian is fitted to the cross-correlation peak.

The HARPS data were reduced by the standard (now public) data reduction software \citep[DRS;][]{Lovis2007}. After reduction, the spectra were cross-correlated with a G2 mask. The resulting cross-correlation function (CCF) is fitted with a Gaussian with its mean as the radial-velocity.

\section{Stellar Properties}\label{sec:spec}
We used the HARPS spectra to determine the stellar atmospheric parameters. The 16 available spectra for J2114-39 were shifted into the laboratory frame and normalised to a continuum level of 1.0. The spectra were then co-added to achieve a combined signal-to-noise ratio, $SNR = 257$.
The following atmospheric parameters for the primary star were derived using {\tt iSpec} \citep{ispec1, ispec2}: the effective temperature, $T_{\rm eff}$, surface gravity, $\log\,g_1$, metallicity, $\rm [Fe/H]$, and microturbulent velocity, $v_{\rm mic}$. We applied the curve-of-growth equivalent widths method with both  the \texttt{WIDTH} \citep{width} and \texttt{MOOG} \citep{sneden2012} radiative transfer codes separately. Solar parameters were input, using those outlined by \citet{ispec2}. The \texttt{SPECTRUM} \citep{spectrum_ll} line list was used  in addition to the \texttt{ATLAS} \citep{atlas} set of model atmospheres. To reduce systematic error, the final stellar atmospheric parameters displayed in Table \ref{tab:derived_params} are weighted averages of the results from both the {\tt WIDTH} and {\tt MOOG} code.

The primary's mass and radius are derived by interpolation of MIST isochrones \citep{Dotter16,Choi16} based on the atmospheric parameters $\rm T_{\rm eff}$, $\rm [Fe/H]$, together with the parallax and infrared colours as input parameters and using a nested sampling approach implemented in the {\tt isochrones} package \citep{Morton15}. The stated uncertainty is the average of the errors calculated from the 16 and 84 percentiles of the resulting distribution. All values are reported in Table~\ref{tab:derived_params}.

\section{Global Modelling}\label{sec:model}

To analyse the data from J2114-39 we use {\tt allesfitter} \citep{allesfitter-paper, allesfitter-code} to perform a simultaneous Markov Chain Monte Carlo (MCMC) modelling of the primary and secondary stars. \texttt{allesfitter} amalgamates many useful \textsc{python} packages frequently used in modelling stellar or planetary systems. At the heart of \texttt {allesfitter} is \texttt{ellc} \citep{ellc} which generates lightcurves and \texttt{celerite} \citep{celerite} for any modelling with Gaussian processes (GPs). To obtain most likely parameters, two types of samplers are used; a nested sampler \citep[\texttt{dynesty;}][]{dynesty} and an affine-invariant MCMC sampler \citep[\texttt{emcee}; ][]{emcee,Goodman2010}.
Here, as in other EBLM papers we use MCMC, since it is less computationally intensive and we have no requirement for model comparison. The nested sampling approach gives similar results to those from the MCMC.
All results presented in the tables and in Section~\ref{sec:conclusion} are from a MCMC fit.

Orbital eccentricity, $e$, in \texttt{allesfitter} is reparameterised with respect to the argument of periastron, $\omega$, as $f_{\rm c} = \sqrt{e}\cos{\omega}$ and $f_{\rm s} = \sqrt{e}\sin{\omega}$ \citep[as in][]{Triaud2011}. For limb darkening, we apply the quadratic law with $T_{\rm eff}$, $\log{g_1}$ and $\rm{[Fe/H]}$ as stellar properties, and adopt the output from \texttt{PyLDTk} \citep{Pyldtk} \citep[using the PHOENIX stellar atmosphere library; ][]{phoenix_lib_pyldtk}. Limb darkening coefficients are reparameterised to $q_1$ and $q_2$ following \citet{Kipping_limbdark} for the fitting process.
All priors are described in Table~\ref{tab:prior_info}. 

From a visual inspection it is clear that there is a level of variability in the {\it TESS} photometry, likely caused by stellar activity. We first tried to apply polynomial and spline functions, but none returned a good fit. To account for intrinsic variability of the star and instrumental noise we fit a GP to the out-of-eclipse photometry assuming a Matérn $3/2$ kernel which detrends short and long term fluctuations. We fit for two hyper-parameters; the amplitude scale $\sigma$ and the length scale $\rho$, as required for this choice of kernel. For the radial-velocity data we add a jitter term (as part of $\ln\sigma_{\rm RV}$) in quadrature with the instrumental white noise error to account for any stellar variability effects, as well as normalised scaling parameters (included in $\ln{\sigma_\mathrm{phot}}$) for the photometry. Other parameters within the model include; the ratio of radii ($R_2/R_1$), inverse scaled semi-major axis ($(R_1+R_2)/a$), cosine of the orbital inclination ($\cos i$), eclipse epoch ($T_0$), orbital period ($P$), radial-velocity semi-amplitude for the primary star ($K_1$), and constant baseline offsets for the radial-velocity instruments ($\Delta F_{\rm RV}$). 

Prior to executing the MCMC sampling, we first perform a visual inspection of the photometric and spectroscopic data to ascertain suitable initial values and priors for the walkers such that the initial fit produces an acceptable and sensible result. When initial values are far from the solutions, walkers take significantly longer to converge, increasing the computational time needlessly for entirely comparable results. For similar reasons, we perform a short MCMC on the radial-velocity data only, before carrying out any joint sampling with the photometry. This allows for more informed estimates of the orbital period and eccentricity of the system. 
Once satisfied with the fit, the prior space is widened to ensure biases were not introduced to the fit by limiting its explorable parameter space.

For the final analysis the MCMC has 60 walkers with 30,000 steps, including 8000 burn-in steps which we discard. The final solutions, considered to be most probable, are determined by the median value of each fitted parameter's posterior distribution where the quoted upper and lower estimates of uncertainty represent the 16/84 percentiles. We consider the chains to have converged to a solution as they were at least 100 times the autocorrelation length for each parameter as well as the chains visually converging in trial plots.

All fitted parameters included in the fit are presented in Table \ref{tab:prior_info} alongside further details on the type of priors used and the selected bounds for the sampling. Physical parameters, derived from the fitted parameters are found in Table \ref{tab:derived_params}.

\renewcommand{\arraystretch}{1.3}
\begin{table*}
    \centering
    \caption{Priors used in the fit and resulting output for the fitted parameters. Uniform priors are marked as $\mathcal{U}$(lower limit, upper limit) and normal priors marked as $\mathcal{N}$(mean, standard deviation). All times are given in $\rm BJD_{UTC}$. The quoted upper limits correspond to the 99.7 percentile of the posterior distribution.}
    \begin{tabular}{l|l|l|l}
        \hline
        \hline
                                          & {\it Parameter} & {\it Prior} & {\it Fit value}  {\it \& Uncertainty} \\
        \hline
        Ratio of radii                   ; & $R_2/R_1$ & $\mathcal{U}(0.080, 0.12)$ & $0.0997 \pm0.0008$ \\
        Inverse scaled semi-major axis  ; & $(R_1 + R_2)/a$ & $\mathcal{U}(0.020, 0.035)$ & $0.02570_{-0.00030}^{+0.00040}$ \\
        Orbital inclination             ; & $\cos i$ &  $\mathcal{U}(0.00, 0.02)$ & $0.0026_{-0.0018}^{+0.0021}$ \\
        Eclipse epoch                   ; & $T_0$ (BJD) & $\mathcal{U}(2\,458\,333.62, 2458333.72) $ & $2\,458\,333.6695_{-0.0012}^{+0.0014}$ \\
        Orbital period                          ; & $P$ (days) & $\mathcal{U}(44.91, 44.95) $ & $44.92471\pm0.00025$ \\
        Radial-velocity semi-amplitude                  ; & $K$ ($\rm{km}s^{-1}$) & $\mathcal{U}(6.0, 7.0)$ & $6.7870_{-0.0069}^{+0.0067}$\\
        Eccentricity and argument of periastron transformation ; & $f\rm{_c}$ & $\mathcal{U}(0.50, 0.60)$ & $0.55915 \pm0.00058$ \\
        Eccentricity and argument of periastron transformation ; & $f\rm{_s}$ & $\mathcal{U}(-0.54, -0.44)$ & $-0.50465 \pm0.00077$\\
        \hline
        \multicolumn{4}{c}{\it Baselines per instrument} \\
        \hline 

        Mat\'ern 3/2 hyperparameter, amplitude scale     ;& $\ln\sigma$ & $\mathcal{U}(-10.0, 0.0)$ & $-7.64_{-0.17}^{+0.21}$ \\
        Mat\'ern 3/2 hyperparameter, length scale     ;& $\ln\rho$ & $\mathcal{U}(-5.0, 5.0)$ & $0.35_{-0.21}^{+0.23}$ \\
        Baseline offset HARPS           ;& $\gamma_{\rm HARPS}$ ($\rm{km}s^{-1}$) & $\mathcal{U}(-25.5, -24.5)$ & $-24.9828\pm0.0036$\\
        Baseline offset FEROS           ;& $\gamma_{\rm FEROS}$ ($\rm{km}s^{-1}$) & $\mathcal{U}(-25.5, -24.5)$ & $-24.9978\pm0.0031$\\
        \hline
        \multicolumn{4}{c}{\it Errors per instrument} \\
        \hline 

        Error scaling for TESS photometry & $\ln \sigma_{\rm TESS}$ & $\mathcal{U}(-10.0, -7.0)$ & $-7.979\pm0.021$\\
        HARPS jitter term               ;& $\ln \sigma_{\rm HARPS}{ (\rm km\,s^{-1})}$ & $\mathcal{U}(-10.0, -2.0)$ & $<-4.6\pm1.6$\\
        FEROS jitter term               ;& $\ln \sigma_{\rm FEROS}{ (\rm km\,s^{-1})}$ & $\mathcal{U}(-10.0, -2.0)$ & $<-3.9$\\
        \hline
        \multicolumn{4}{c}{\it Limb Darkening Coefficients} \\
        \hline 
        Transformed limb darkening ; &$q_{\rm 1;TESS}$ & $\mathcal{N}(0.1545856, 0.5)$& $0.172_{-0.073}^{+0.11}$\\
        Transformed limb darkening ; &$q_{\rm 2;TESS}$ & $\mathcal{N}(0.520898, 0.5)$ &  $0.47_{-0.20}^{+0.27}$\\
        \hline
    \end{tabular}
    \label{tab:prior_info}
\end{table*}

\section{Parameter derivation}\label{sec:adjustment}

To extract accurate parameters on the secondary star, we need accurate parameters on the primary. Thanks to {\it Gaia} DR3 \citep{Gaia2016} the stellar radius ($R_1$) is better estimated than the mass ($M_1$) that traditionally relies on applying stellar models. However, with our lightcurve modelling we can constrain $M_1$ using the stellar density. From Kepler's law, stellar density ($\rho_{\rm 1}$) is defined as
\begin{equation}\label{eq:rho}
    \frac{M_{\rm 1}}{R_{\rm 1}^3} = \frac{4\pi ^2}{G\,P^2} \left( \frac{a}{R_{\rm 1}} \right)^3 - \frac{M_{\rm 2}}{R_{\rm 1}^3},
\end{equation}
where $G$ is the gravitational constant, and $a$ is the semi-major axis \citep{EBLMI}.

In planetary cases the second term ($M_2/R_1^3$) is typically considered negligible. However, for binary stars, it must be included, since its contribution is no longer insignificant. For eclipsing binaries we are able to determine the primary and secondary masses independent of any assumption of $M_1$.

We start by rearranging Eqn.\ref{eq:rho} for $M_2$ and define the total mass term as
\begin{equation}
    M = \frac{4\pi ^2}{G\,P^2} \left( \frac{a}{R_{\rm 1}} \right)^3 R_1^3 = M_1 + M_2.
\end{equation}
Then using the observables fit via the method described in Section~\ref{sec:model} from each step in the MCMC we calculate the mass function, $f_m$ \citep{Hilditch2001},
\begin{equation}\label{eq:fm_obs}
    f_m = \frac{(1-e^2)^{3/2}}{\sin^3{i}} \,\frac{P\,K_1^3}{2\pi \,G} .
\end{equation}
As the mass function can also be expressed as
\begin{equation}\label{eq:fm}
    f_m = \frac{M_{\rm 2}^3}{(M_{\rm 1} + M_{\rm 2})^2} = C,
\end{equation}
we can substitute in our expression from the stellar density equation to calculate a secondary mass\begin{equation}\label{eq:m2}
    M_2 = \sqrt[3]{CM^2}.
\end{equation}
Finally with a value for $M_2$ we can now use Eqn.\ref{eq:rho} to calculate the primary mass as
\begin{equation}\label{eq:m1}
    M_1 = M - M_2.
\end{equation}

From the eclipse signal analysis, ${a}/{R_{\rm 1}}$ is a modelled variable along with all other observables in Eqn \ref{eq:rho} and \ref{eq:fm_obs} which allows for the stellar masses to be calculated. To ensure the error in the the primary radius, $R_1$ (the only assumed value from outside the modelling), is reflected in the final mass uncertainty we assign a normal Gaussian distribution and draw random samples to be worked through in the same manner as all other variables.

\section{Results and Conclusion}\label{sec:conclusion}

After the global modelling and parameter derivation, we determine a mass function \massfunction, indicating a low-mass companion. Then we calculate the secondary's surface gravity \loggsec, which is found in absolute terms \citep{Southworth2007} from parameters fit with the MCMC. This value alone confirms the secondary is a dense star, near the bottom of the main-sequence. Assuming parameters for the primary star (as in Section~\ref{sec:spec}; $R_1 = 1.283 \pm0.012~\,\rm R_\odot$ and Section~\ref{sec:adjustment}: $M_1 = 0.998 \pm 0.052~\,\rm M_\odot$), we find the secondary stellar companion to be a late M-dwarf with a mass \mass\, and a radius \radius\, and thus a mass ratio \massratio\, with the primary. If we recalculate the surface gravity of the primary star using the obtained mass and radius, we find that the value is consistent within 1 sigma with the spectroscopic surface gravity, regardless of the choice of primary mass value. From the fit, we also find the system has a clearly detected orbital eccentricity \eccentricity, which is not unusual for a binary with this period \citep[e.g.][]{Triaud2017}.

\begin{figure}
    \centering
    \includegraphics[width=0.49\textwidth]{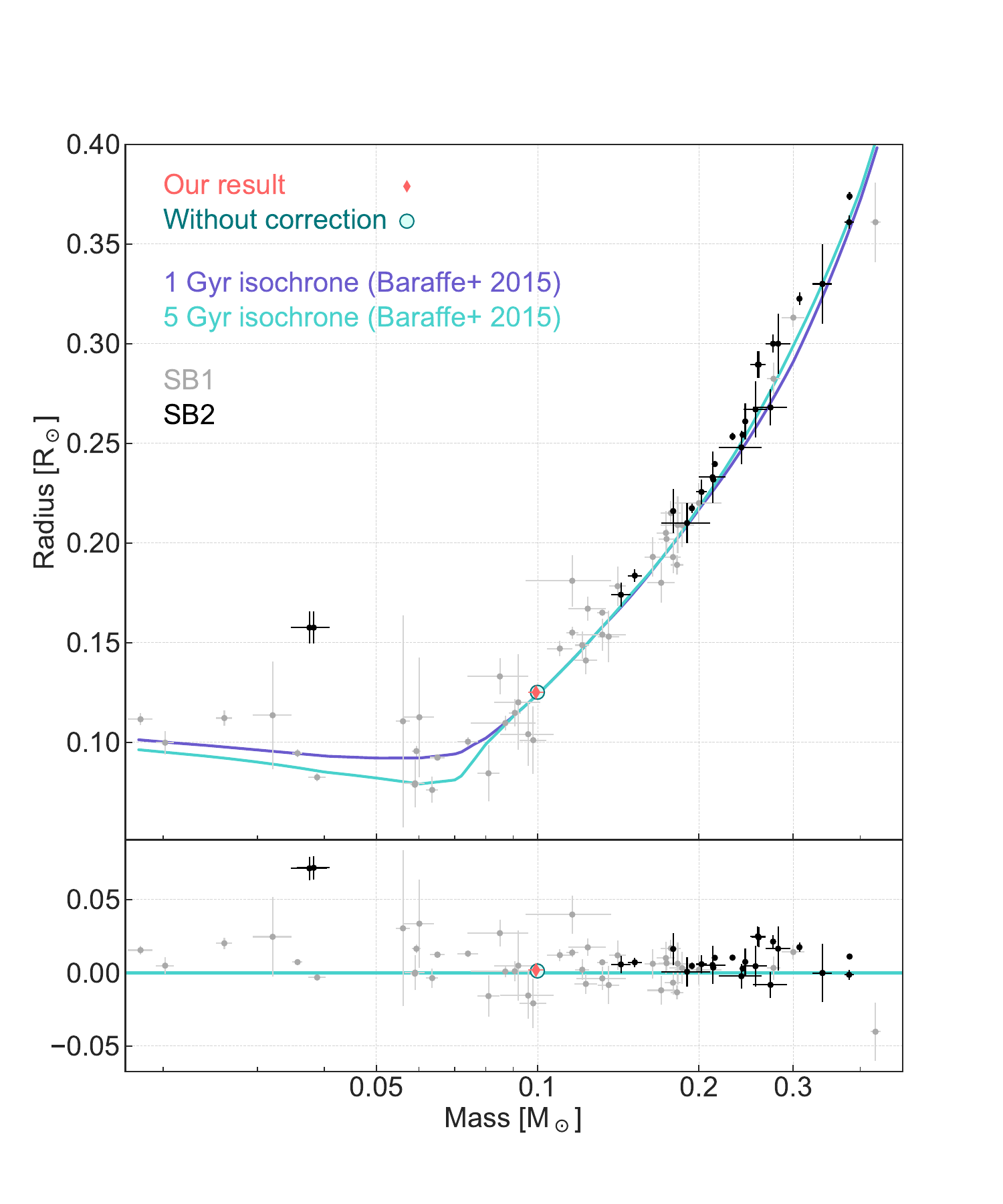}
    \caption{
    Mass-radius diagram with logarithmic mass axis, showing published single-lined systems (SB1) in grey, double-lined binaries (SB2) in black and EBLM~J2114-39 results overlapping in turquoise and coral colour. The $1~\rm Gyr$ and $5~\rm Gyr$ isochrones are shown in purple and teal respectively. The lower panel highlights differences between the objects and the $5~\rm Gyr$ isochrone. Black and grey data points from compendium (and cited papers) in \citet{Triaud2020}, with additional points from \citet{morales2009}. 
    }
    \label{fig:mass_radius_plot}

\end{figure}

Without following the parameter derivation procedure described in Section \ref{sec:adjustment}, we would have calculated the mass of the secondary star through equation \ref{eq:fm} where we would have assumed the denominator to be $M_1^2$ as commonly done when $M_2 \ll M_1$. By using the primary mass value calculated from isochrones, we would then have found the mass of the secondary companion to be $M_2 = 0.0999~\pm 0.0045~\,\rm M_\odot$ instead of \mass. Although these results are consistent within error margins (as can be seen in Fig. \ref{fig:mass_radius_plot}) this represents a $\sim 1.0\%$ systematic bias towards higher masses (and would therefore bias the overall population towards less inflated objects). Appendix \ref{sec:app_error_cal} details why the fractional uncertainty on mass is smaller for the secondary star than for the primary.

EBLM~J2114-39b has a mass approaching the hydrogen burning limit, which is a key region for empirical calibrations of the mass-radius relation. As illustrated in the mass-radius plot (Fig. \ref{fig:mass_radius_plot}), our target is consistent within $1~\sigma$ with the \citet{baraffe2015} models continuing a trend observed in \citet{Boetticher2019} and \citet{2023Swayne}, where fully convective secondary stars do not appear to be systematically inflated whereas higher mass M-dwarfs are. Several mechanisms have been proposed to explain radius inflation, however these explanations do not predict why fully convective stars would be different and not show inflation. Here we speculate that this might have to do with convective overshooting, which is known to affect modelled stellar parameters \citep[e.g.][]{Baraffe2023}. For a fully convective star such as EBLM~J2114-39\,B, convective overshooting is essentially irrelevant since there is no penetration of convective materials into a radiative interior. However slightly more massive stars, with a small radiative core, might be very sensitive to how convective overshooting is modelled, since any plume of convective material into the small radiative interior would likely affect a greater fraction of the core, and might produce large differences in stellar evolution. 

Another aspect that makes J2114-39B stand out in this context is its long orbital period. Typically other objects of a similar mass are found in binary systems with periods $< 10~\rm\,days$. As such EBLM~J2114-39B is likely not much influenced by its primary.

\renewcommand{\arraystretch}{1.3}
\begin{table}
    \caption{Parameters of EBLM~J2114-39. $^{(*)}$ from Sec. \ref{sec:spec}, and $^{(**)}$ from GAIA DR3. Primary mass values given from the MIST isochrones $M_{1, \rm iso}$, as well as from the method described in Section \ref{sec:adjustment}. The quoted upper limits correspond to the 99.7 percentile of the posterior distribution.}
    \centering
    \begin{tabular}{l|l|l}
          & $ $ \\
         \hline
            \hline
            \multicolumn{3}{c}{{{\it Primary star parameters}}} \\
            \hline
            Primary mass$^*$ & $M_{1,\rm iso}$ ($\mathrm{M_{\odot}}$) & $1.003 \pm 0.042$  \\
            Primary mass & $M_{1}$ ($\mathrm{M_{\odot}}$) & $0.998_{-0.054}^{+0.049}$  \\
            Primary radius$^*$ & $R_1$ ($\rm R_\odot$)& $1.283 \pm 0.012$ \\
            Effective temperature$^*$ & $T_{\rm eff}$ (K) & $5667 \pm 113$  \\
            Metallicity$^*$ & [Fe/H] (dex)& $-0.001 \pm 0.032$ \\
            Spectroscopic surface gravity$^*$ & $\log g_1$ (cgs)& $3.936	\pm 0.010$  \\
            Microturbulent velocity$^*$ &$v_{\rm mic}$ ($\rm km\,s^{-1}$) & $0.839	\pm 0.010$  \\
            Parallax$^{**}$ & Plx (mas) & $4.30 \pm 0.02$   \\
            \hline
            \multicolumn{3}{c}{{{\it Secondary star derived parameters}}} \\
            \hline
            Scaled primary radius & $R_1/a$ & $0.02338_{-0.00028}^{+0.00036}$ \\
            Scaled secondary radius & $R_2/a$ & $0.002329_{-0.000029}^{+0.000044}$ \\
            Orbital eccentricity & $e$ & $0.56733_{-0.00026}^{+0.00027}$ \\
            Argument of periastron & $\omega$ (deg) & $317.933_{0.073}^{0.071}$\\
            Secondary mass & $M_2$ ($\mathrm{M_{\odot}}$) & $0.0993_{-0.0035}^{+0.0031}$ \\
            Secondary radius & $R_2$ ($\rm R_\odot$)& $0.1250 \pm 0.0016$\\
            Mass Ratio & $q$ & $0.0995_{-0.0017}^{+0.0020}$ \\
            Impact parameter & $b$ & $<0.120$\\
            Total eclipse duration & $T_\mathrm{tot}$ (h) & $11.624_{-0.097}^{+0.108}$ \\
            Full eclipse duration & $T_\mathrm{full}$ (h) & $9.469_{-0.097}^{+0.109}$ \\
            Eclipse depth & $\delta_\mathrm{undil}$ & $0.01141 \pm -0.00017$\\
            Limb darkening & $u_1$ & $0.39_{-0.11}^{+0.10}$\\
            Limb darkening & $u_2$ & $0.03_{-0.18}^{+0.21}$\\
           \hline
           Mass function& $f_m$ ($\rm M_\odot$)& $0.0008127\pm 0.0000028$ \\
           Companion surface gravity & $\log g_2$ (cgs)& $5.222_{-0.016}^{+0.011} $ \\
           \hline
    \end{tabular}
    \label{tab:derived_params}
\end{table}

\section*{Acknowledgements}
We thank the staff at ESO's observatory of La Silla for their kind attention, and the extra work they produced during the COVID pandemic, while travel restrictions prevented us from going to the telescope. We also thank the anonymous reviewer whose comments helped to improve the quality of the paper.

RB acknowledges support from FONDECYT project 11200751 and additional support from ANID -- Millennium  Science  Initiative -- ICN12\_009.

A.J. acknowledges support from ANID -- Millennium  Science  Initiative -- ICN12\_009 and  from FONDECYT project 1210718.

M.H. acknowledges support from ANID - Millennium Science Initiative - ICN12\_009.

MRS acknowledges support from the UK Science and Technology Facilities Council (ST/T000295/1), and the European Space Agency as an ESA Research Fellow.

T.T. acknowledges support by the DFG Research Unit FOR 2544 "Blue Planets around Red Stars" project No. KU 3625/2-1.
T.T. further acknowledges support by the BNSF program "VIHREN-2021" project No. KP-06-DV/5.

This research was supported by UK Science and Technology Facilities Council (STFC) research grant number ST/M001040/1.

This research is supported from the European Research Council (ERC) under the European Union's Horizon 2020 research and innovation programme (grant agreement n$^\circ$ 803193/BEBOP), and by a Leverhulme Trust Research Project Grant (n$^\circ$ RPG-2018-418).


\section*{Data Availability}
All HARPS and FEROS data can be obtained from the ESO archive. All {\it TESS} data can be obtained from MAST.
 



\bibliographystyle{mnras}
\bibliography{references} 




\appendix

\section{Error calculation for stellar masses} \label{sec:app_error_cal}

The fractional error on the secondary star is smaller than the fractional error on the primary mass. This may be surprising, but is actually a direct result of the fit to the RVs being very good and the different powers in the mass ratio of Eqn \ref{eq:fm}. 
From Eqn \ref{eq:m2}, we can write that 
$$M_2^3 = C\,M^2.$$
Using the rules of error propagation, we can show that 
$$\left(\frac{\sigma_{M_2}}{M_2}\right)^2 = \frac{4}{9} \left(\frac{\sigma_{M}}{M}\right)^2 + \frac{1}{9} \left(\frac{\sigma_{C}}{C}\right)^2.$$
The error in $M$ can be calculated from Eqn \ref{eq:m1}:
$$\sigma_{M}^2 = \sigma_{M_1}^2 - 
\sigma_{M_2}^2$$
after which it follows that 
$$\left(\frac{\sigma_{M_2}}{M_2}\right)^2 = \frac{4}{9} \frac{\sigma_{M_1}^2 - \sigma_{M_2}^2}{\left(M_1+M_2\right)^2} + \frac{1}{9} \left(\frac{\sigma_{C}}{C}\right)^2.$$
Rearranging this equation finally gives us 
$$\left(\frac{\sigma_{M_2}}{M_2}\right)^2 = \frac{4\,M_1^2}{9\,(M_1+M_2)^2 + 4\,M_2^2} \left(\frac{\sigma_{M_1}}{M_1}\right)^2 $$
$$\qquad\qquad+ \frac{(M_1+M_2)^2}{9\,(M_1+M_2)^2 + 4\,M_2^2} \left(\frac{\sigma_{C}}{C}\right)^2$$
In the limiting case of a perfect fit to the RVs and thus $\sigma_C=0$, this reduces to

$$\left(\frac{\sigma_{M_2}}{M_2}\right)^2 = \frac{4\,M_1^2}{9\,(M_1+M_2)^2 + 4\,M_2^2} \left(\frac{\sigma_{M_1}}{M_1}\right)^2 < \left(\frac{\sigma_{M_1}}{M_1}\right)^2$$

where it is now obvious that the precision on $M_2$ is better than the one on $M_1$. This will still hold for very good fits to the RVs where the fractional error in $C$ remains small, as is the case for our binary system. At some point, the error from the RV fit will be large enough that it is no longer the case, for example in the regime of planets orbiting stars where the RV fits are not as well constrained. We note that with a slight adjustment, it can also be shown that the secondary mass precision could also be better than the primary mass precision in the case of non-transiting binary systems where the primary mass comes from isochrone analysis. In that case $\sigma_{M}^2 = \sigma_{M_1}^2 + \sigma_{M_2}^2$.

\section{RV data} \label{sec:app_rv_data}

\begin{table}
    \caption{HARPS observations of J2114-39. All times are given in $\rm BJD_{UTC}$.}
    \centering
    \begin{tabular}{c|c|c|c|c}
       BJD (day) & RV $(\rm km s^{-1})$ & $\sigma$ $(\rm km s^{-1})$ & Exposure Time (s) & SNR \\
       \hline
       2459171.599696 & -28.74679 & 0.02187 &  300 & 6.7\\
       2459425.784454 & -27.12687 & 0.01073 &  300 &10.6\\
       2459433.716703 & -28.52944 & 0.01232 &  300 & 9.5\\
       2459462.782953 & -24.95927 & 0.01488 &  300 & 8.5\\
       2459479.607512 & -28.61051 & 0.02043 &  300 & 6.1\\
       2459516.622836 & -27.34166 & 0.01040 &  300 & 11.7\\
       2459519.599470 & -27.93342 & 0.01410 &  300 & 8.7\\
       2459535.500919 & -25.31486 & 0.00703 &  300 & 15.4\\
       2459703.926928 & -28.60003 & 0.00585 &  300 & 18.1\\
       2459737.860021 & -26.58939 & 0.00350 &  900 & 27.9\\
       2459761.691474 & -20.98407 & 0.00583 &  900 & 18.3\\
       2459823.663107 & -25.47391 & 0.00667 &  900 & 16.7\\
       2459828.499860 & -26.78018 & 0.00444 &  900 & 24.1\\
       2459841.604636 & -28.86199 & 0.00446 &  900 & 23.2\\
       2459874.628686 & -27.04324 & 0.00460 &  900 & 23.3\\
       2459897.565604 & -17.22338 & 0.00511 &  900 & 21.3\\

    \end{tabular}
    \label{tab:harps_obvs}
\end{table}

\begin{table}
    \centering
    \caption{FEROS observations of J2114-39. All times are given in $\rm BJD_{UTC}$.}
    \begin{tabular}{c|c|c|c}
       BJD (day) & RV $(\rm km s^{-1})$ & $\sigma$ $(\rm km s^{-1})$ & SNR \\
       \hline
        2458738.55034 & -22.4797 & 0.0072 & 43.3\\
        2458742.71029 & -24.5180 & 0.0074 & 40.0\\
        2458782.52082 & -21.8770 & 0.0069 & 45.7\\
        2458793.56281 & -26.3889 & 0.0061 & 52.5\\
        2458813.54052 & -28.3868 & 0.0071 & 40.5\\
        2458815.52633 & -27.0573 & 0.0062 & 51.1\\
    \end{tabular}
    \label{tab:feros_obvs}
\end{table}



\bsp	
\label{lastpage}
\end{document}